\documentclass[final]{aipproc}

\layoutstyle{6x9}

\usepackage{amsmath}
\usepackage{graphicx}
\usepackage{bm}
\usepackage{here}
\usepackage[dvips,monochrome]{color}

\newcommand{\bra}[1]{\langle #1|}
\newcommand{\ket}[1]{|#1\rangle}

\begin{document}

\title{Coherent spectral weights of Gutzwiller-projected superconductors}

\author{Samuel Bieri}{
  address={Institute of Theoretical Physics, Ecole Polytechnique
  F\'{e}d\'{e}rale de Lausanne (EPFL), CH-1015 Lausanne, Switzerland}
}

\author{Dmitri Ivanov}{}

\pacs{71.10.Li,74.72.-h,71.18.+y,71.10.Fd}

\keywords{HTSC, lattice fermions, strongly correlated systems,
unconventional superconductivity}


\begin{abstract}
We analyze the electronic Green's functions in the superconducting
ground state of the $t$-$J$ model using Gutzwiller-projected wave
functions, and compare them to the conventional BCS form. Some of
the properties of the BCS state are preserved by the projection: the
total spectral weight is continuous around the quasiparticle node
and approximately constant along the Fermi surface. On the other
hand, the overall spectral weight is reduced by the projection with
a momentum-dependent renormalization, and the projection produces
electron-hole asymmetry in renormalization of the electron and hole
spectral weights. The latter asymmetry leads to the bending of the
effective Fermi surface which we define as the locus of equal
electron and hole spectral weight.
\end{abstract}

\maketitle

\section{Introduction}

High temperature superconductivity (HTSC) is one of the most
intriguing phenomena in modern solid state physics. Experimentally,
HTSC is observed in layered cuprate compounds. The undoped cuprates
are antiferromagnetically ordered insulators which develop the
characteristic superconducting ``dome'' upon doping with charge
carriers.



HTSC is interesting not only for promising technological
applications, but also from a theoretical point of view. The
relevant ingredients for HTSC are believed to be the following.
\begin{itemize}
\item Low dimensionality (2d).
\item Strong short-range repulsion between electrons.
\item Doped Mott insulator.
\end{itemize}
Taking these 3 ingredients, HTSC is modeled in the tight-binding
description by large-$U$ Hubbard models or $t$-$J$ models on the
square lattice:
\begin{equation}\label{eq:tJ}
H_{tJ} = -t\sum_{\langle i,j\rangle} P_G\, c_{i\sigma}^\dagger
c_{j\sigma}\, P_G + J \sum_{\langle i,j\rangle} ({\bm S}_i\cdot{\bm
S}_j - \frac{n_i n_j}{4})
\end{equation}
where $n = c_{\sigma}^\dagger c_{\sigma}$, ${\bm S} = \frac{1}{2}
c_{\sigma}^\dagger\bm{\sigma}_{\sigma\sigma'}c_{\sigma'}$ and
$\bm{\sigma}$ are the Pauli matrices.\footnote{Repeated indices are
summed over.} The Gutzwiller projector $P_G = \Pi_i
(1-n_{i\uparrow}n_{i\downarrow})$ prevents electrons from occupying
the same lattice site.

The non-perturbative nature of the $t$-$J$ model makes it an
outstanding problem to solve in dimensions larger than one.
Analytical techniques (renormalized or slave-boson mean-field
theories \cite{rmft,kotliar88}) are very crude and numerical
techniques (e.g.\ exact diagonalization or cluster DMFT) are
restricted to very small clusters or infinite dimensions, or they
fail on the sign problem (QMC). An alternative approach was
suggested by Anderson shortly after the experimental discovery of
HTSC, when he proposed a Gutzwiller-projected BCS wave function as
superconducting ground state for cuprates \cite{anderson87}.
Following this conjecture, many variational studies have been
performed on the basis of what is called Anderson's (long range) RVB
state. This state turned out to have very low variational energy,
close to exact ground state energies, as well as high overlap with
the true ground states of small $t$-$J$ clusters
\cite{gros88,hasegawa89}. On the other hand, many experimental facts
about cuprate superconductors can be reproduced and are consistent
with the variational results: e.g.\ clearly favored $d$-wave pairing
symmetry, doping dependency of the nodal Fermi velocity and the
nodal quasiparticle weight. Many of these successful efforts
following Anderson's proposal are summarized in the ``plain vanilla
RVB theory'' of HTSC, recently reviewed in \cite{anderson04}.

With help of the relatively recent technique of angle-resolved
photoemission spectroscopy (ARPES), experimentalists can probe the
electronic structure of low-lying excitations inside the copper
planes. The intensity measured in ARPES is proportional to the
one--particle electronic spectral function: $I_{PES}({\bm k},
\omega)\; \propto A( {\bm k}, \omega) $ \cite{ARPES}. It is
therefore interesting to explore spectral properties within the
framework of Gutzwiller-projected variational quasiparticle (QP)
excitations.

In this contribution we will discuss some of our results reported in
\cite{bieri06}. For more details, in particular for more reference
to experimental studies, we invite the reader to consult that paper.

\section{Coherent spectral weights}

Anderson's RVB state is given by
\begin{equation}\label{eq:H}
\ket{H} \propto P_H P_G \ket{d\text{BCS}(\Delta,\mu)}\; .
\end{equation}
We further define projected BCS quasiparticle excitations in a
similar way,
\begin{equation}\label{eq:excit}
\ket{H,\bm{k},\sigma} \propto P_H P_G
\gamma^\dagger_{\bm{k}\sigma}\ket{d\text{BCS}}\; .
\end{equation}
The unprojected states in Eqs.~\eqref{eq:H} and \eqref{eq:excit} are
the usual ingredients of the BCS theory, $\ket{d\text{BCS}} =
\Pi_{\bm{k},\sigma}\gamma_{\bm{k}\sigma}\ket{0}$,
$\gamma_{\bm{k}\sigma} = u_{\bm{k}} c_{\bm{k}\sigma} +{\small\sigma}
v_{\bm{k}} c_{-\bm{k}\bar\sigma}^\dagger$, $u_{\bm{k}}^2 =
\frac{1}{2}\left(1-\frac{ \xi_{\bm{k}} }{E_{\bm{k}}}\right) =
1-v_{\bm{k}}^2$, $E_{\bm{k}} = \sqrt{\xi_{\bm{k}}^2 +
\Delta_{\bm{k}}^2}$, $\xi_{\bm{k}} = - 2 [\cos(k_x) + \cos(k_y)] -
\mu$, $\Delta_{\bm{k}} = \Delta[\cos(k_x) - \cos(k_y)]$. $P_G$ is
the Gutzwiller projector and $P_H$ projects on the subspace with $H$
holes. $\ket{H}$ and $\ket{H,\bm{k},\sigma}$ are normalized to one.
The wave functions have two variational parameters, $\Delta$ and
$\mu$, which we adjusted to minimize the energy of the $t$-$J$
Hamiltonian \eqref{eq:tJ} for the experimentally relevant value
$J=0.3$ and every doping level. Note that in the RVB theory,
$\Delta$ and $\mu$ are variational parameters without direct
physical significance; physical quantities like excitation gap,
superconducting order, or chemical potential must be calculated
explicitly.

The spectral weights of the coherent low-lying quasiparticles
\eqref{eq:excit} can be written as
\begin{subequations}\label{eq:weights}\begin{eqnarray}
Z^{+}_{\bm k} =& |\bra{H - 1, \bm k,\sigma} c_{{\bm k},\sigma}^{\dagger} \ket{H}|^2\\
Z^{-}_{\bm k} =& |\bra{H + 1, \bm k,\sigma} c_{-{\bm k},\bar\sigma}
\ket{H}|^2 \; .\end{eqnarray}\end{subequations} These weights are
measured in ARPES experiments as the residues of the spectral
function $A({\bm k}, \omega)$ \cite{ARPES}. Note that in
conventional BCS-theory, $Z_{\bf k}^{+} = u_{\bm k}^2$ and $Z_{\bf
k}^{-} = v_{\bm k}^2$.

\section{Method: VMC}

The variational Monte Carlo technique (VMC) allows to evaluate
fermionic expectation values of the form $\bra{\psi} O \ket{\psi}$
for a given state $\ket{\psi}$. In order to calculate the spectral
weights \eqref{eq:weights} by VMC, the following exact relations can
be used.
\begin{subequations}\label{eq:z}\begin{eqnarray}\label{eq:z1}
Z^{+}_{\bm k} &=& \frac{1+x}{2} -  \langle c^{\dagger}_{\bm k\sigma}
c_{\bm k\sigma}\rangle \;,\\
\label{eq:z2} Z^{+}_{\bm k} Z^{-}_{\bm k} &=& \lvert \bra{H+1}
c_{\bm k\uparrow} c_{-\bm k\downarrow} \ket{H-1}\rvert^2 \; ,
\end{eqnarray}\end{subequations} where $x$ is hole doping
\cite{yunoki05,bieri06}.

We use VMC to calculate the superconducting order parameter
$\Phi_{\bm k} = \lvert \langle c_{\bm k\uparrow} c_{-\bm
k\downarrow} \rangle\rvert$, as well as diagonal matrix elements in
the optimized $t$-$J$ ground state \eqref{eq:H}. Using relations
\eqref{eq:z}, we can then derive the spectral weights
\eqref{eq:weights}. The disadvantage of this procedure is large
errorbars around the center of the Brillouin zone where both
$Z^{+}_{\bm k}$ and $\Phi_{\bm k}$ are small.

\section{Results}
\begin{figure}[h]
\includegraphics[scale=0.48,angle=0]{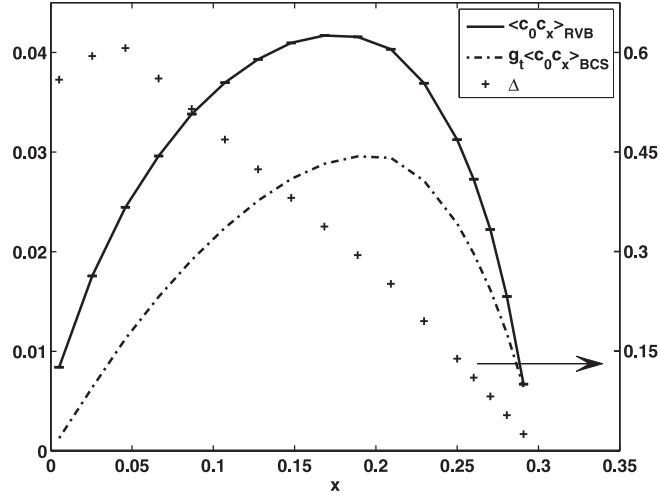}
\caption{Doping dependency of the nearest-neighbor superconducting
order parameter $\Phi_{ij}$ (calculated in the 14$\times14$ system).
The errorbars are smaller than the symbol size. The same quantity
calculated in the Gutzwiller approximation is also shown for
comparison. The variational parameter $\Delta$ is shown with the
scale on the right.\label{fig:cc_x}}
\end{figure}

In Fig.~\ref{fig:cc_x}, we plot the nearest-neighbor superconducting
orderparameter $\langle c_{i \uparrow} c_{j \downarrow} \rangle$ as
a function of doping. The curve shows close quantitative agreement
with the result of Ref.~\cite{paramekanti0103}, where the authors
extracted the same quantity from the long-range asymptotics of the
nearest-neighbor pairing correlator,
$\lim_{r\rightarrow\infty}\langle c_0 c_{\delta} c^{\dagger}_r
c_{r+\delta}^{\dagger}\rangle$. With the method employed here, we
find the same qualitative and quantitative conclusions of previous
authors \cite{gros88,paramekanti0103}: vanishing of
superconductivity at half filling, $x\to 0$, and at the
superconducting transition on the overdoped side, $x_c\simeq 0.3$.
The optimal doping is near $x_{opt}\simeq 0.18$. In the same plot we
also show the commonly used Gutzwiller approximation where the BCS
orderparameter is renormalized by the factor $g_t = \frac{2 x}{1+x}$
\cite{rmft}. The Gutzwiller approximation underestimates the exact
value by approximately $25\%$.

\begin{figure}[!h]
\begin{minipage}[b]{0.5\linewidth} 
\centering
\includegraphics[scale=.44]{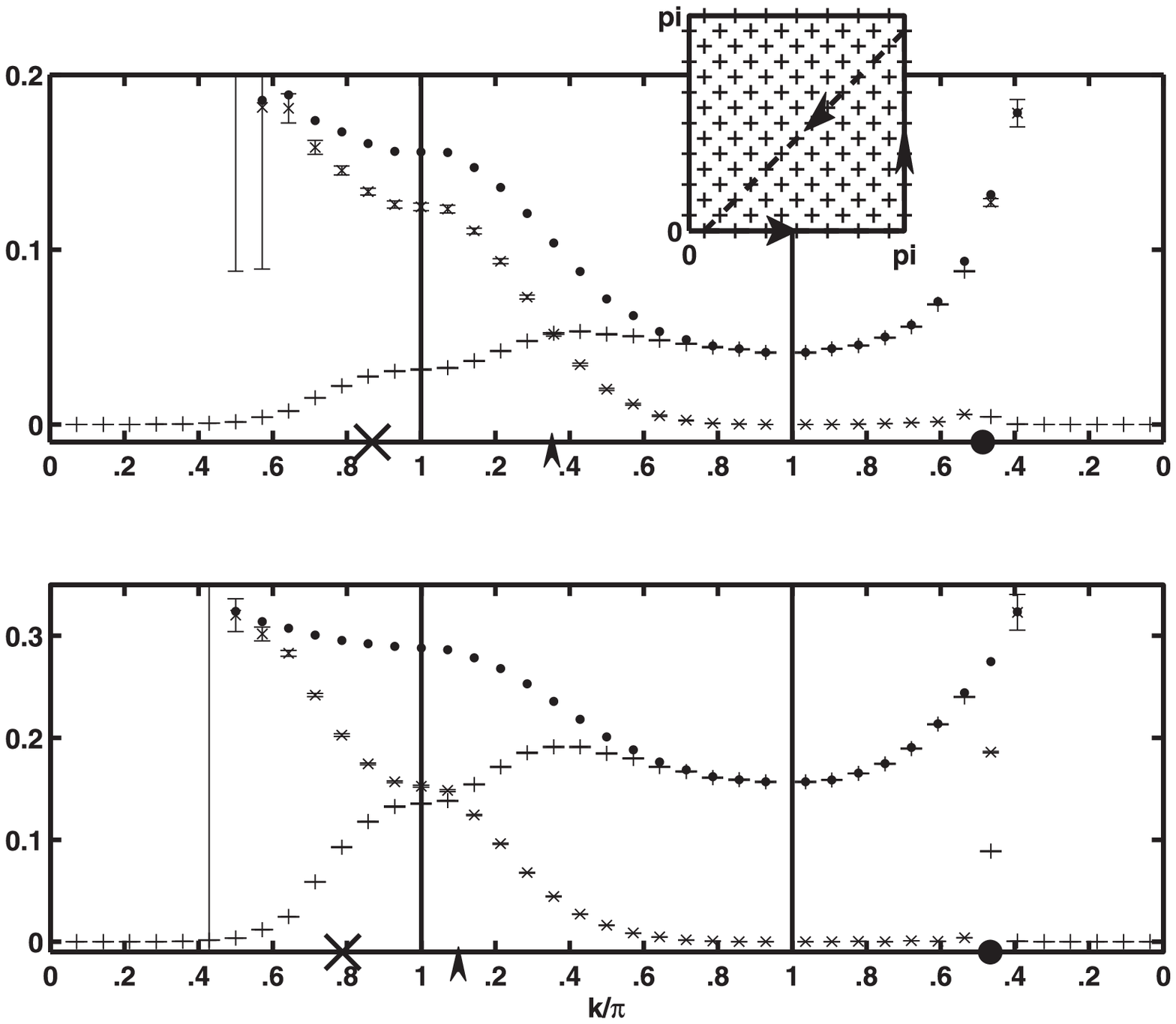}
\end{minipage}
\begin{minipage}[b]{0.5\linewidth} \centering
\includegraphics[scale=.44]{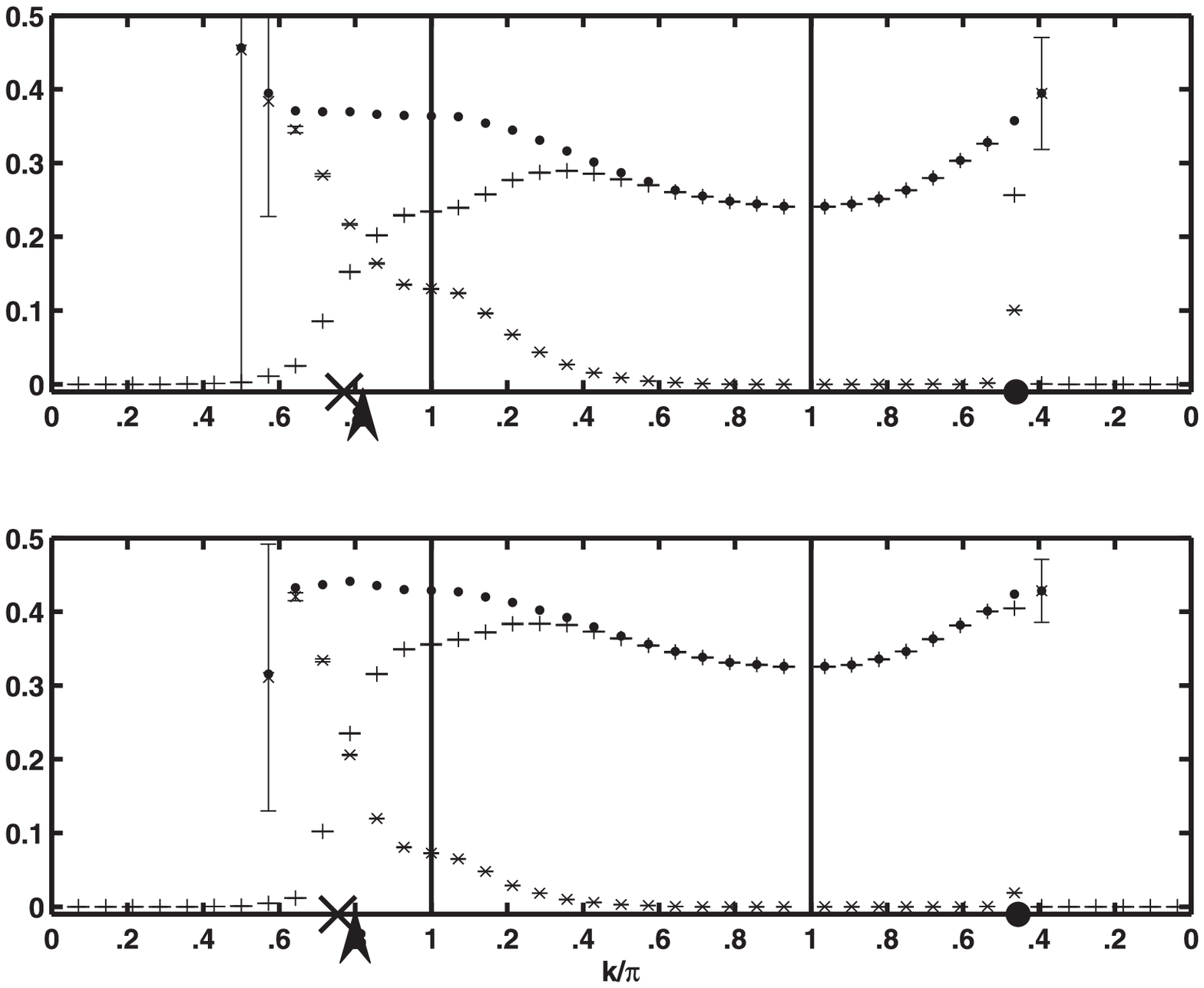}
\end{minipage}
\caption{QP spectral weights for 6 holes (upper left plot, $x\simeq
3\%$), 22 holes (lower left plot, $x\simeq 11\%$), 34 holes (upper
right plot, $x\simeq 17\%$), and 46 holes (lower right plot,
$x\simeq 23\%$) on 196 sites. The spectral weights are plotted along
the contour $0\to(0,\pi)\to(\pi,\pi)\to 0$ (shown in inset). Plus
signs ($+$) denote the spectral weight $Z_{\bm k}^{+}$, crosses
($\times$) denote $Z_{\bm k}^{-}$, errorbars are shown. Solid dots
denote their sum, the total spectral weight $Z_{\bm k}^{tot}$,
errorbars not shown. On the horizontal axis, the star ($*$) denotes
the intersection with the unprojected Fermi surface along the
$0\to(0,\pi)$ direction; the thick dot is the nodal point. Both
$Z_{\bm k}^{+}$ and $Z_{\bm k}^{-}$ jump at the nodal point, while
$Z_{\bm k}^{tot}$ is continuous. The intersection with the effective
Fermi surface (see text) is marked by an arrowhead. On the diagonal
(last segment), $k/\pi$ is given in units of $\sqrt{2}$.
\label{fig:cut_z_196}}
\end{figure}

\begin{figure}[!h]
\includegraphics[scale=.9,angle=0]{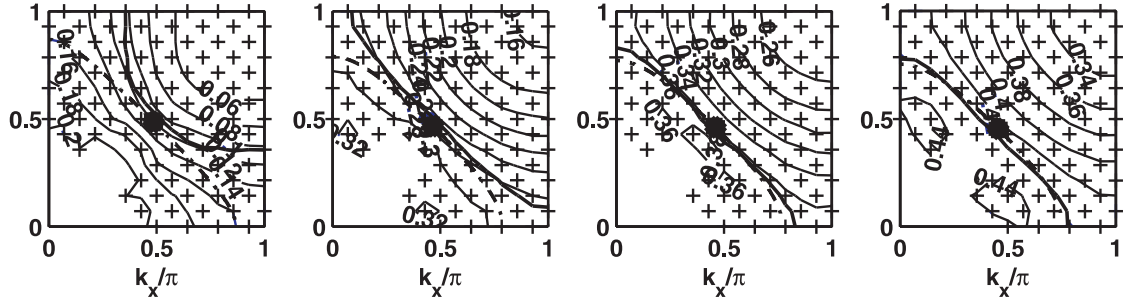}
\caption{Contour plots of the total QP spectral weight $Z_{\bm
k}^{tot}$. The effective FS (full line) and unprojected FS (dashed
line) are also shown. The doping levels are $x\simeq 3\%$, $11\%$,
$17\%$, and $23\%$ (from left to right). The $+$ signs indicate
points where the values are known within small errorbars (see
Section \textit{Method: VMC}).\label{fig:nFS}}
\end{figure}

In Fig.~\ref{fig:cut_z_196}, we plot the spectral weights $Z^+_{\bm
k}$, $Z^{-}_{\bm k}$, and $Z^{tot}_{\bm k}$ along the contour
$0\to(0,\pi)\to(\pi,\pi)\to 0$ in the Brillouin zone for different
doping levels. Figure~\ref{fig:nFS} shows the contour plots of
$Z^{tot}_{\bm k}$ in the region of the Brillouin zone where our
method produces small errorbars. From these data, we can make the
following observations.
\begin{itemize}
\item In the case of an unprojected BCS wave function, the total
spectral weight is constant and unity over the Brillouin zone.
Introducing the projection operator, we see that for low doping
($x\simeq3\%$), the spectral weight is reduced by a factor up to
$20$. The renormalization is asymmetric in the sense that the
electronic spectral weight $Z^{+}_{\bm k}$ is more reduced than the
hole spectral weight $Z^{-}_{\bm k}$. For higher doping ($x\simeq
23\%$), the spectral weight reduction is much smaller and the
electron-hole asymmetry decreases.
\item Since there is no electron-hole mixing along the zone
diagonal ($d$-wave), the spectral weights $Z^{+}_{\bm k}$ and
$Z^{-}_{\bm k}$ have a discontinuity at the nodal point. Our data
shows that the total spectral weight is continuous across the nodal
point. Strong correlation does not affect these features of
uncorrelated BCS-theory.
\end{itemize}



\subsection{Effective Fermi surface} \label{sec:FS}
In strongly interacting Fermi systems, the notion of a Fermi surface
(FS) is not clear at all. There are, however, several experimental
definitions of the FS. Most commonly, ${\bm k}_F$ is determined in
ARPES experiments as the maximum of $|{\bm \nabla}_{\bm k}n_{\bm
k}|$ or the locus of minimal gap along some cut in the $\bm
k$-plane. The theoretically better defined locus of $n_{\bm k} =
1/2$ is also sometimes used. The various definitions of the FS
usually agree within the experimental uncertainties \cite{ARPES}.
The different definitions of the FS in HTSC were recently analyzed
theoretically in Refs.~\cite{gros06, sensarma06}.

Here, we propose an alternative definition of the Fermi surface
based on the ground state equal-time Green's functions. In the
unprojected BCS state, the underlying FS is determined by the
condition $|u_{\bm k}|^2 = |v_{\bm k}|^2$. We will refer to this as
the \textit{unprojected FS}. Since $|u_{\bm k}|^2$ and $|v_{\bm
k}|^2$ are the residues of the QP poles in the BCS theory, it is
natural to replace them in the interacting case by $Z^{+}_{\bm k}$
and $Z^{-}_{\bm k}$, respectively. We will therefore define the
\textit{effective FS} as the locus $Z^{+}_{\bm k} = Z^{-}_{\bm k}$.

In Fig.~\ref{fig:nFS}, we plot the unprojected and the effective FS
which we obtained from VMC calculations. The contour plot of the
total QP weight is also shown. It is interesting to note the
following points.
\begin{itemize}
\item In the underdoped region, the effective FS is open and bent
outwards (hole-like FS). In the overdoped region, the effective FS
closes and embraces more and more the unprojected one as doping is
increased (electron-like FS).
\item Luttinger's rule \cite{luttinger61} is clearly violated in the
underdoped region, i.e.\ the area enclosed by the effective FS is
not conserved by the interaction; it is larger than that of the
unprojected FS.
\item  In the optimally doped and overdoped region, the total spectral weight
is approximately constant along the effective FS within errorbars.
In the highly underdoped region, we observe a small concentration of
the spectral weight around the nodal point ($\simeq 20\%$).
\end{itemize}
Large ``hole-like'' FS in underdoped cuprates has also been reported
in ARPES experiments by several groups
\cite{ino04,yoshida03,shen03}.

It should be noted that a negative next-nearest hopping $t'$ would
lead to a similar FS curvature as we find in the underdoped region.
We would like to emphasize that our original $t$-$J$ Hamiltonian as
well as the variational states do not contain any $t'$. Our results
show that the outward curvature of the FS is due to strong
repulsion, without need of $t'$. The next-nearest hopping terms in
the microscopic description of the cuprates may not be necessary to
explain the FS topology found in ARPES experiments. Remarkably, if
the next-nearest hopping $t'$ is included in the variational ansatz
(and not in the original $t$-$J$ Hamiltonian), a finite and negative
$t'$ is generated, as it was shown in Ref.~\cite{himeda00}.
Apparently, in this case the unprojected FS has the tendency to
adjust to the effective FS. A similar bending of the FS was also
reported in the recent analysis of the current carried by
Gutzwiller-projected QPs \cite{nave06}. A high-temperature expansion
of the momentum distribution function $n_{\bm k}$ of the $t$-$J$
model was done in Ref.~\cite{putikka98} where the authors find a
violation of Luttinger's rule and a negative curvature of the FS.
Our findings provide further evidence in this direction.

A natural question is the role of superconductivity in the
unconventional bending of the FS. In the limit $\Delta\to 0$, the
variational states are Gutzwiller-projected excitations of the Fermi
sea and the spectral weights are step-functions at the (unprojected)
FS. In a recent paper \cite{yang06}, it was shown that $\lim_{{\bm
k}\rightarrow {\bm k}^{+}_F} Z^{+}_{\bm k} = \lim_{{\bm
k}\rightarrow {\bm k}^{-}_F} Z^{-}_{\bm k}$ for the projected
Fermi-sea, which means that the unprojected and the effective FS
coincide in that case. This suggests that the ``hole-like'' FS
results from a non-trivial interplay between strong correlation and
superconductivity. At the moment, we lack a qualitative explanation
of this effect. However, it may be a consequence of the proximity of
the system to the non-superconducting ``staggered-flux'' state
\cite{lee00,ivanov03} or to antiferromagnetism near half-filling
\cite{paramekanti0103,ivanov06}.

\section{Acknowledgement}
SB would like to thank the organizers of the \textit{Eleventh
Training Course in the Physics of Correlated Electron Systems and
HTSC} in Salerno, Italy, where this work was presented. We would
like to thank George Jackeli for many illuminating discussions and
continuous support. We also thank Claudius Gros, Patrick Lee, and
Seiji Yunoki for interesting discussions. This work was supported by
the Swiss National Science Foundation.

\end{document}